\documentclass[aps,twocolumn,prl,showpacs,color,psfig,epsf]{revtex4-1}
\usepackage{amsmath}
\usepackage{color}
\usepackage{amsfonts}
\usepackage{epsf}
\usepackage{graphicx}
\usepackage{natbib}

\usepackage{ulem}
\definecolor{red}{rgb}{1,0,0}
\definecolor{blue}{rgb}{0,0,1}
\definecolor{darkgreen}{RGB}{3,101,3}
\definecolor{antonio}{RGB}{51,153,255}
\definecolor{francesco}{RGB}{255,0,255}

\begin{document}

\title{Large fluctuations and dynamic phase transition in a system of self-propelled particles}

\author{F. Cagnetta$^1$, F. Corberi$^2$,  G. Gonnella$^1$, A. Suma$^3$}
\affiliation{$^1$ Dipartimento di Fisica, Universit\`{a}  di Bari, {\rm and} Sezione INFN di Bari, via Amendola 173,
70126 Bari, Italy \\
$^2$Dipartimento di Fisica âE.R.Caianielloâ {\rm and} INFN, Gruppo Collegato di Salerno, {\rm and} CNISM, Unit\`{a} di Salerno, Universit\`{a} di Salerno, via Giovanni Paolo II 132, 8408 Fisciano (SA),
Italy \\
$^3$ SISSA â Scuola Internazionale Superiore di Studi Avanzati, Via Bonomea 265, 34136 Trieste, Italy \\
}

\begin{abstract}
We study the statistics, in stationary conditions,  of the work $W_\tau$ done
by the active force in different systems of self-propelled particles in a time $\tau$. 
We show the existence of a critical value
$W_\tau ^\dag$ such that fluctuations with $W_\tau >W_\tau ^\dag$ correspond to
configurations where interaction between particles plays a minor role whereas those
with $W_\tau < W_\tau ^\dag$ represent states with single particles dragged
by clusters. This two-fold behavior is fully mirrored by the probability distribution 
$P(W_\tau )$ of the work, which does not obey the large-deviation principle for $W_\tau <W_\tau ^\dag$.
This pattern of behavior can be interpreted as due to a phase transition occurring at the level of
fluctuating quantities and an order parameter is correspondingly identified. 

\end{abstract}


\maketitle

In equilibrium systems the deviations of an observable quantity from the average
occur with a probability regulated by the Boltzmann-Einstein expression 
$\exp{\{\Delta S/k_B\}}$~\cite{landauv5}, where $k_B$ is the Boltzmann constant
and $\Delta S$ is the entropy increase due to such a fluctuation.

In dynamical contexts one is often confronted with the related problem of finding the probability 
distribution of certain observables measured over a time interval $\tau$.
A sound  mathematical framework for a general description of fluctuations, 
which can be also applied to far-from-equilibrium systems, is provided by the large deviation theory~\cite{touchette2009aa}.
When a {\it large deviation principle} (LDP) holds, the probability distribution 
$P(W_\tau)$ of a given quantity $W_\tau$ is characterised by a rate function 
$I(W_\tau)=- \lim _{\tau \to \infty}\frac{1}{\tau}  \ln P(W_\tau)$.
General predictions are in some cases available for $I(W_\tau)$, for example in diffusive
models ~\cite{bertini2001aa,*bertini2005aa,bodineau2004aa,*bodineau2005pre},
\textcolor{black}{where $W_\tau$ is the particle current flowing in systems in contact with
two reservoirs at different densities.}
Probability distributions exhibiting a non-analytical behavior interpretable as a  phase transition~\cite{bertini2010aa,bodineau2004aa,hurtado2011aa,bunin2012aa,harris2005jstat,szavitnossan2014aa,gradenigo2013aa,gambassi2012prl,filiasi2014jstat,*corberi2015large,zannetti2014ab,*corberi2013aa} have been encountered, recently attracting a considerable interest.

A possibility to test and extend the above ideas in a new far-reaching context is offered by active matter.
The inherently far from equilibrium systems belonging to this class, either biological or artificial in nature, display a number of nontrivial properties without analogue in passive, equilibrium materials~\cite{Marchetti13,Elgeti15,gonn15}.
A suspension of self-propelled particles, for instance, may phase separate into a dense and a gaseous phase, even in the absence of any attractive interaction~\cite{Tailleur08,fily2012athermal,Redner13,Stenhammer13,Buttinoni13b,Suma14,fodor2016far}. 
Furthermore, active particles accumulate at boundaries~\cite{Elgeti09,*Elgeti13}, follow 
in the dilute limit a Boltzmann profile with an effective temperature
when sedimenting~\cite{Palacci10,cugl-mossa1,*cugl-mossa2,ginot2015nonequilibrium}, etc.
\textcolor{black}{Addressing the properties of fluctuations occurring at a mesoscopic level in these
systems is fundamental for a full characterisation of their functions, as, for instance, in the case of molecular motors \cite{astumian1998fluctuation}.
In the context of active brownian motion,} 
large deviations have been studied 
in experiments  with an 
asymmetric particle interacting with a vibrated granular medium~\cite{kumar2011aa,*kumar2015aa}.
\textcolor{black}{By considering the fluctuations of a quantity akin to 
the work defined in Eq.~(\ref{defw}), one can test a fluctuation 
relation~\cite{hurtado2011ab,ganguly2013stochastic,chaudhuri2014active} which quantifies the relative 
probability of small-scale entropy consuming events that go beyond the second principle.}

\textcolor{black}{In this Letter, we study fluctuations in different systems of interacting 
active brownian particles propelled by a force directed along their polar axis 
\cite{peruani06,Suma14,Redner13,fily2012athermal}.} 
Specifically, we will consider the probability $P(W_{\tau})$ of the work done by the 
active force in a time interval $\tau$ on each particle.
Our results show that, while for values of $W_\tau$ larger than a critical threshold $W_\tau^\dag$ the 
LDP holds, it fails for $W_\tau <W_\tau ^\dag$ because the rate function 
$I(W_\tau)$ vanishes; in this sector $\ln P(W_{\tau})\propto -W_\tau$ behaves linearly.
Such a twofold behavior can be discussed in terms of a transition between a phase, for 
$W_\tau >W_\tau ^\dag$, where the particles are basically free and one, for 
$W_\tau <W_\tau ^\dag$, where they can be dragged by moving clusters.
Correspondingly, an order parameter related to the relative orientation between a 
particle and the direction of motion of the surrounding aggregate can be defined. 
\textcolor{black}{These results hold true for the different types of particles studied, 
pointing towards a general character.} 

\textcolor{black}{We study models consisting of  $N$ particles with different shapes 
(for more model details,  parameters used 
and simulation methods see the supplementary material (SM)\footnote{The Supplemental Material is provided at [URL will be inserted by publisher] and includes additional
Refs.~\cite{Cicotti,vankampen,seifert2005aa,evans1993aa,redner2001guide}}), either spherical colloids, dumbbells, 
{\it tetrabells}, or convex rod-like molecules (see Fig. SM5).
For concreteness we describe below the case of dumbbells,} 
a classical example of anisotropic particles which has been 
considered in many active matter studies~\cite{valeriani2011colloids}.
\textcolor{black}{Results for other kinds of particles 
will be presented in SM.}

Dumbbells are made up by two beads, a head and a tail, both having diameter $\sigma$. These 
are held together by a finitely extensible nonlinear elastic (FENE) spring.  
Any pair of beads interact via the purely repulsive  
Weeks-Chandler-Anderson (WCA) potential~\cite{Weeks},
namely a Lennard-Jones interaction truncated at its minimum. 
Denoting with $U$ the full potential (including both WCA and FENE terms), the evolution of the position ${\mathbf {x}_i}$ of the $i$-th bead is given by a Langevin equation, 
\begin{equation}
m\frac{d^2 {\mathbf x}_i}{dt^2}=-\gamma\frac{d{\mathbf x}_i}{dt}-\nabla_i U+
{\boldsymbol F_a}+
\sqrt{2k_BT\gamma} {\boldsymbol \xi}_i(t), 
\label{bd}
\end{equation}
where $\gamma$ is the friction,  
$\nabla_i=\frac{\partial}{\partial {\mathbf {x}_i}}$,
$T$ is the temperature of a thermal bath in contact with the system, 
$m$  is the mass, ${\boldsymbol F_a}$ 
is a tail-head-directed active force 
with constant magnitude $F_a$ ,
and ${\boldsymbol \xi}_i(t)$ is an uncorrelated Gaussian noise with zero mean and unit variance.
We study a two-dimensional system.
 We set the parameters~
 \footnote{All physical quantities are expressed in reduced units of mass $m$, energy $
 \epsilon$ and length $\sigma$ (the last two related to the WCA potential), which are set equal 
 to one. The time unit is the standard Lennard-Jones time $\tau_{LJ}=\sigma\sqrt\frac{m}{\epsilon}$. 
Other important simulation parameters, in reduced units, are $\gamma=10$, $k_BT=0.01$ and we set 
$k_B=1$. The simulation box is a square with side length $200\sigma$ or more. 
 For more informations on the potential parametrization and other simulation constants, see SM.} 
as to have a strongly overdamped dynamics,  which is realistic for 
microswimmers; the stiffness of the FENE springs is also 
strong enough~\cite{Note2} that the distance between the head and the 
tail is in practice constant and equal to $\sigma$. 
Dimensionless numbers relevant for the following are the area fraction covered by
the particles, $\phi=\frac{N\pi\sigma^2}{2A}$, where $A$ is the area of the 
simulation domain, and  the P\'eclet number  ${\rm Pe}=\frac{2F_a\sigma}{k_BT}$ \cite{Suma14}.

The phase diagram and other properties of this active dumbbell system have been studied in~\cite{Suma14,Suma14b,Suma14c,*Suma15a,joyeux2016pressure}.
When  $\phi$ exceeds a P\'eclet dependent threshold,
an initial homogeneous state phase separates~\cite{Suma13}. 
On the other hand, for sufficiently low  values of $\phi$,
particles form    
small aggregates that  do not coalesce. 
\textcolor{black}{This is the situation that we find in all our simulations.
In some cases, as specified 
in the caption of  Fig.~\ref{fig2},  
the system was reported~\cite{Suma14,Suma13} to be slightly inside the binodal line, suggesting
that macroscopic aggregation could be observed on much longer times than those 
addressed in this paper.
}  



\begin{figure}[!h]
\begin{center}
  \begin{tabular}{cc}
       \includegraphics[width=1\columnwidth]{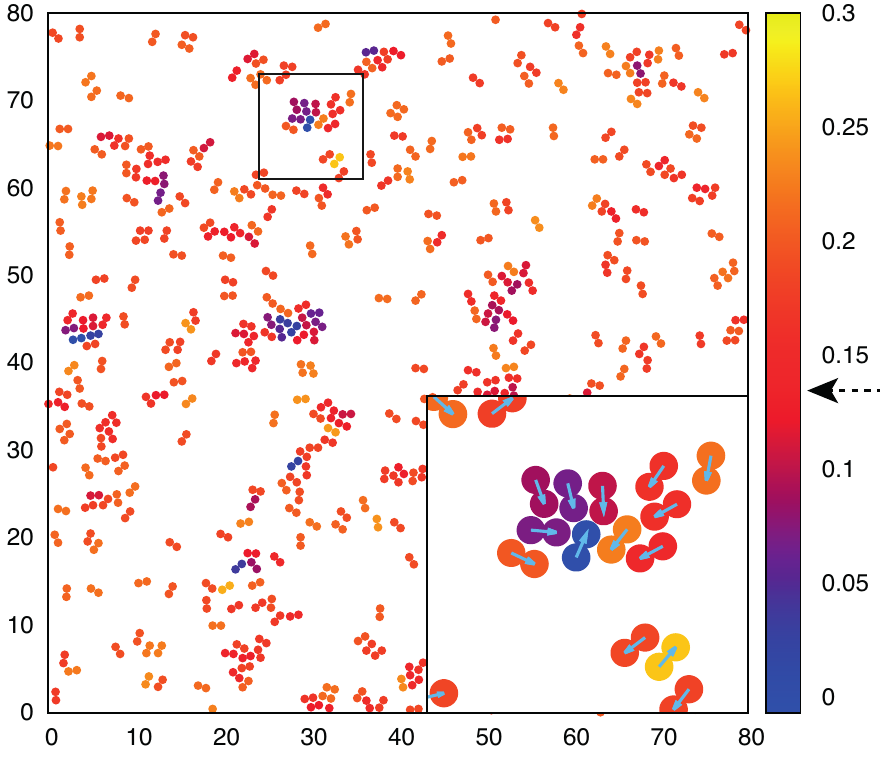}\\
    \end{tabular}
\caption{A typical stationary configuration at $\phi=0.1$ and ${\rm Pe}=200$, \textcolor{black}{just inside the binodal line~\cite{Suma14},} 
 \textcolor{black}{(only a portion of the system of size $L=200$ is shown)}. For each dumbbell, the color represents the value of $ W_{\tau}$ defined in Eq. (\ref{defw}) 
(with $\tau=10$)
according to the color-code in the bar on the right.
The black arrow indicates the value of $W^\dag (\tau)$.
The inset is a magnification of the highlighted box, 
where also the arrows representing the active force directions have been drawn
(see movies M1\textcolor{black}{-M5} in SM to visualize the evolution of the system
\textcolor{black}{(for dumbbells, colloids, tetrabells, and rods of different aspect ratio, respectively))}.}

\label{fig1}
\end{center}
\end{figure}

An instance of the kind of configurations we work with is shown in Fig.~1.
 One observes 
groups of dumbbells travelling together due to steric effects
and aggregates rarely exceeding ten units. 
For this system, under  stationary conditions \textcolor{black}{(see SM, Sec. IA)}, we evaluate
the observable
\begin{equation}W_{\tau} = \frac{2}{\tau} \int_t^{t + \tau}{ \bold F_a(t') \cdot \bold v_i(t')\, dt'} 
\label{defw}
\end{equation}
representing, for each dumbbell, 
the  work (per unit time) done by the  active force. 
Being proportional to the time averaged projection 
of the centre of mass velocity $\bold v_i$
along the main direction of the dumbbell, this quantity is akin to that measured in the experiments mentioned above \cite{kumar2011aa},
and it also represents the entropy production for individual particles (see SM).
We call such an observable {\it active work}, and the main object of our study is the probability $P(W_\tau )$ of its outcomes.
This can be evaluated analytically only for a single non-interacting particle (see SM for details).
Denoting it as $P_0(W_\tau)$, this distribution turns out to be gaussian with 
average $<W_{\tau}>_0 = 2F_a^2/\gamma$ and variance $<W^2_\tau>_0= \frac{4F_a^2k_BT}{\tau\gamma}$.
It is shown in Fig.\ref{fig2} 
(continuous black curve) where, in order to have a better representation,
we plot $\sqrt{<W^2_\tau>_0} \,P(W_{\tau})$ {\it vs} 
$(W_{\tau}-<W_{\tau}>_0)/\sqrt{<W^2_\tau>_0}$. 
In the same figure, data obtained by numerical integration of 
Eq. (\ref{bd}) for finite density of particles 
(two values are presented) are also displayed. 
These show that, irrespective of $\phi$, the $W_{\tau}$ distribution becomes gaussian in the small ${\rm Pe}$ limit.
The curves for ${\rm Pe}=1$ are indeed indistinguishable from the analytical ones. 
The same is true for any ${\rm Pe}$, in the limit of very small area fraction 
(see Fig.SM2).
On the other hand, the character of the distribution changes dramatically by 
increasing ${\rm Pe}$ at a fixed finite value of $\phi$. The 
curve remains peaked around a value close to the non-interacting 
one $<W_{\tau}>_0$, is still gaussian on the whole region to the right
and immediately on the left of it, but changes abruptly as to have 
an approximately linear behavior of $\ln P(W_\tau)$ for
$W_\tau$ smaller than a critical threshold $W_\tau ^\dag>0$, represented
by a vertical arrow in Fig. \ref{fig2}.
This feature, which represents the central and new result of this paper, is  clearly
manifest for ${\rm Pe} \gtrsim 50$.  
Similar results are found for the different kind of particles before mentioned, 
and are shown in Fig. SM6.
\textcolor{black}{It is worth mentioning that rods with an aspect ratio similar to the one used in this paper lack a macroscopic phase-separation transition~\cite{Baskaran12}, but display even a more pronounced discontinuity at $W^\dag _\tau$.}
\textcolor{black}{This suggests that the fluctuation phenomenon we observe is not straightforwardly
related to the macroscopic motility induced phase-separation.}

\begin{figure}
\centerline{\includegraphics[width=1\columnwidth]{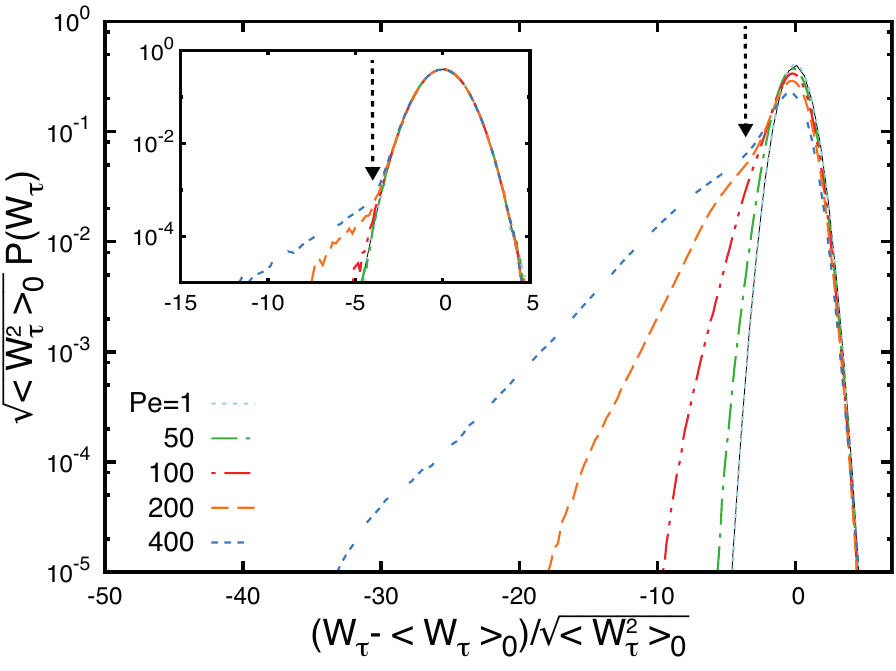}}
\caption{Probability distributions (on log-scale) of $ W_{\tau}$, with $\tau=20$, 
for ${\rm Pe}=1,50,100,200,400$, at $\phi=0.1$ (main) and $\phi=0.001$ (inset). \textcolor{black}{All  cases fall inside the homogeneous low-density phase, except those at  $\phi=0.1$ and ${\rm Pe}=200,400$ that are just inside the co-existence region.}
The threshold $W_\tau^\dag$ is signalled by vertical arrows. The continuous 
black curve is the analitical result 
for a single dumbbell (see SM). 
}
\label{fig2}
\end{figure}

\begin{figure}
\begin{center}
  \begin{tabular}{cc}
       \includegraphics[width=\columnwidth]{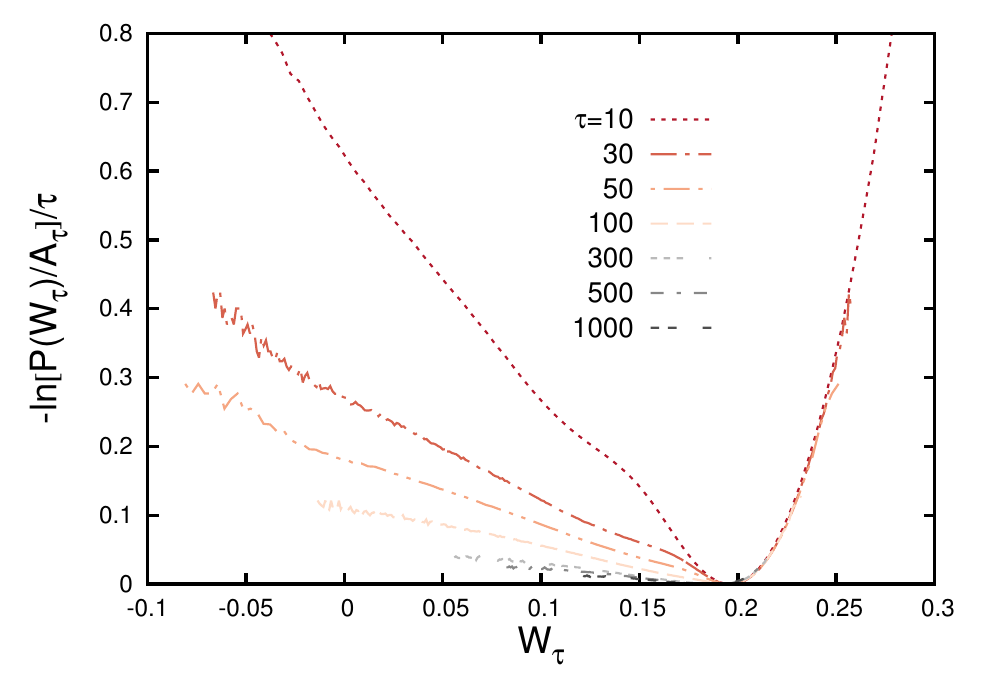}\\
    \end{tabular}
\caption{The quantity  $\frac{1}{\tau} \ln [P(W_{\tau})/ A_{\tau}]$, 
is plotted {\it vs} $W_\tau$ for different
values of $\tau $, ${\rm Pe}=200$ and $\phi=0.1$. } 
\label{fig3}
\end{center}
\end{figure}

Let us now consider the effect of changing
 $\tau$. The linear decay to the left of the maximum is observable from $\tau \sim 5$ 
until $\tau \simeq 1000$. 
This is because for $\tau \lesssim 5$ the distribution resembles more a gaussian 
while for $\tau \gtrsim  1000 $ 
the tail cannot be detected with a 
significative statistics. 
In Fig. \ref{fig3} we plot $ \frac{1}{\tau} \ln [P(W_{\tau})/ A_{\tau}]$ {\it vs}
$W_\tau$ for different choiches of
$\tau \in [10-1000]$, where $A_{\tau}$ is the maximum of $P(W_\tau)$ (this is 
done to better compare the curves at different $\tau$). 
According to the LDP, in such a graph one should observe data collapse
of outcomes with different $\tau$ on a master-curve $I(W_\tau)$ -- the rate function.
This was found in the experiments with vibrated particles \cite{kumar2011aa}.
Instead, what we have is something clearly different. Data collapse 
is only obtained for values of $W_\tau$ larger than
$W_\tau ^\dag$, whereas curves are well separated in the region with $W_\tau < W ^\dag_\tau$.
These results imply that, at least in the range of times accessed in our simulations,
the LDP is obeyed for $W_\tau>W_\tau^\dag$ but not for $W_\tau <W_\tau ^\dag$.
In this region the large-$\tau$ limit  of $(1/\tau)\ln P(W_\tau)$ vanishes
and a different scaling takes place, as discussed in
the SM. \textcolor{black}{This implies that fluctuations are suppressed more softly for large 
$\tau$ with respect to the usual case when the LDP holds.}
We emphasize that this anomalous behavior is not
restricted to particular choices of the model parameters, but is found with the same
characteristics in a whole range of densities and P\'eclet numbers, \textcolor{black}
{and for all kinds of particles considered}. Specifically, we
observe that the breakdown of the LDP is always flanked by the appearence of the 
linear behavior of $\ln P(W_\tau)$ on the left of the maximum.

In order to understand which events do contribute to the linear tails of $\ln P(W_{\tau})$ we 
isolated in simulations particle trajectories with a fixed value of $W_{\tau}$. 
The colours in Fig. \ref{fig1} represent  the distribution of $W_{\tau}$ 
in a specific realization of our system.  
An event with a value of $W_\tau $ much smaller than $W_\tau ^\dag$
(marked by a horizontal sign in the colour palette on the right)
is shown in 
the zoomed part of the figure.
One sees that the particle in the centre 
has its polar axis  pointing  against a cluster moving in the opposite direction. 
The blue dumbbell is dragged by the cluster against its active force, resulting in a value of $W_{\tau}$ significantly smaller than the average.
\textcolor{black}{Once the relevant mechanism is identified, a simple kinetic argument can be developed to infer the existence of a threshold $W^\dag _\tau$ and estimate its dependence on the model parameters.
This is discussed in Sec. VI of SM. It turns out that 
$W^\dag _\tau \propto (F_a^2/\gamma ) [1-c/(D_R\tau)]$,
where $c$ depends on the kind of active particle considered and $D_R$ is
the rotational diffusion coefficient defined in SM. This dependence
has been confirmed in our simulations for all kinds of particles considered (see
Fig. SM8).}

Identifying the mechanism producing $W^\dag _\tau$ 
does not clarify how it originates the linear behavior of $\ln P(W_\tau)$.
Actually, the probabilities in Fig. \ref{fig2} resemble very closely those 
found analytically in \textcolor{black}{reference statistical mechanics models,
such as the Gaussian model or the zero-range process \cite{zannetti2014ab,harris2005jstat,szavitnossan2014aa}.
Here non-analiticities have been discovered, and deviations corresponding
to the the linear tail of $\ln P$ have been linked to 
a condensation transition taking place in the space of fluctuations.
The system concentrates in a 
narrow region of phase-space, similarly to what happens when a gas turns into a liquid
or in Bose-Einstein condensates.} Something very similar is found also in the present active matter system. 
Indeed, while for $W_\tau >W_\tau ^\dag$ the velocity of each dumbbell is symmetrically 
distributed
around the average (as in the single particle case) and can take any possible orientation --
thus filling 
the whole phase-space,
when $W_\tau <W_\tau ^\dag$ this orientational symmetry is broken,
since such velocity is set to that of the surrounding cluster.

Following this train of thought we argue that a possible order parameter should be related to the relative 
orientation between the dumbbell and its surroundings. Thus, calling ${\boldsymbol R_i}$ the total force 
felt by the $i$-th dumbbell due to interactions with other particles, and $\theta_i$ the angle between its 
main axis and ${\boldsymbol R_i}$, we define a microscopic, instantaneous order parameter $\tilde{m}(t)$, 
which equals $-\cos{\theta_i}$ when the dumbbell at hand is in contact with the others (let us recall 
that the WCA potential is truncated at distances of order $\sigma$, see SM), otherwise it is null.
The overall order parameter $m(W_{\tau})$ is obtained by averaging $\tilde{m}(t)$ over all the histories of 
time length $\tau$ which result in an active work $W_{\tau}$.

\begin{figure}
\centerline{\includegraphics[width=1\columnwidth]{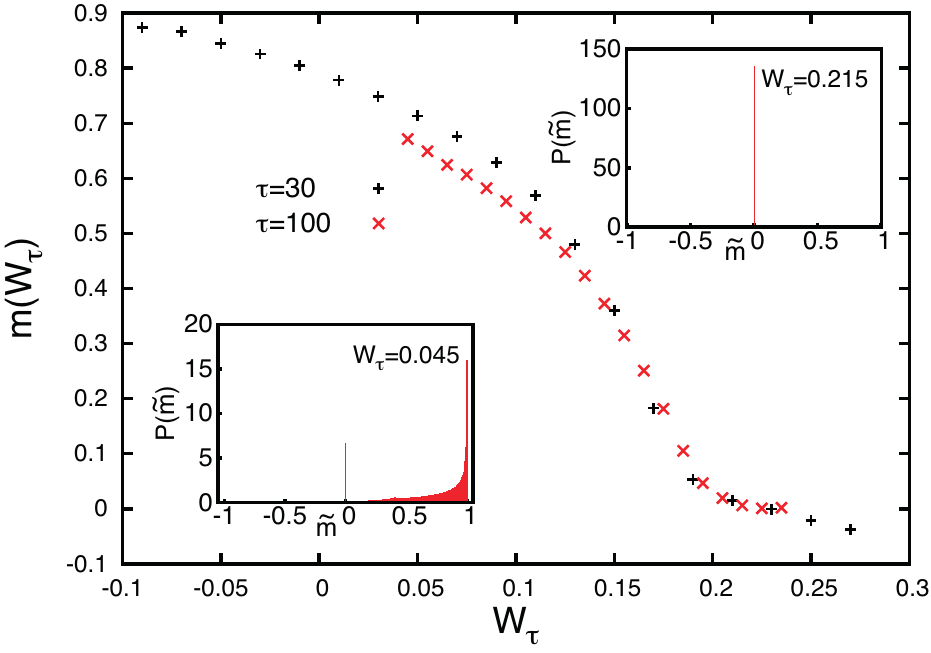}}
\caption{The quantity $m(W_\tau $) is plotted {\it vs} $W_\tau $ for $\tau =30$ and $\tau =100$
(see key), with ${\rm Pe}=200$ and $\phi=0.1$.  In the two insets the histogram of the values of $\tilde m$ is shown 
(for a case with $W_\tau =0.215 >W_\tau ^\dag$ on the right and with 
$W_\tau = 0.045 <W_\tau ^\dag$
on the left. }
\label{fig4}
\end{figure}

Fig.~\ref{fig4}  shows the behavior of  $m(W_\tau)$ as a function of $W_{\tau}$.
One sees that  $m$ is zero for $W_\tau \gtrsim 0.2$, a value that we identify with  
$W^\dag _\tau$ obtained from Fig. \ref{fig2}, while it increases for 
$W_{\tau} \le W^\dag _\tau$ and tends to 1 at large negative $W_\tau$.
The reason is that the instantaneous parameter $\tilde{m}(t)$ equals $1$ when the dumbbell under analysis is being pulled against its active force direction.
Fig.~\ref{fig4}
clearly demonstrates that such a mechanism is effective below $W^\dag _\tau$ --
making $m$ finite and positive --
and becomes progressively more important as $W_\tau$ is further lowered.
This behaviour is robust as $\tau $ is changed, as it is shown in Fig.~\ref{fig4}.
This is exactly the kind of property one would expect for an 
order parameter,
with $W_\tau$ playing the role of an external control parameter (akin to temperature),
and $W_\tau ^\dag$ that of a critical point separating a broken-symmetry phase (here for $W_\tau <W_\tau ^\dag$)
from a symmetric one (for $W_\tau >W_\tau ^\dag$).
Not only the average $m$ has the behavior expected
for an order parameter, but also its fluctuating value $\tilde m$.
The distribution of its values, shown in the insets of Fig.~\ref{fig4}, displays indeed a single 
sharp peak centered around $\tilde m=0$ for $W_\tau \ge W_\tau ^\dag$ while it develops, as soon as $W_\tau ^\dag$ is crossed, an additional peak at $\tilde m=1$ whose height grows as $W_{\tau}$ decreases,
analogously to what occurs in usual equilibrium phase transitions.
Here the height of the peak around $\tilde m=1$ represents the fraction of 
$\tau$ for which the dumbbell has been pulled backwards by a cluster 
\footnote{It is worth noticing that $\tilde m$ 
is strictly related to $ W_{\tau} $. In fact, as shown in SM, 
interactions between the specified dumbell and the surrounding ones enter the quantity 
${\boldsymbol F_a} \cdot {\mathbf v_i}$ of Eq. (\ref{defw}) through the term
$ [1/(2\gamma)] {\boldsymbol F_a} \cdot {\boldsymbol R_i}$, and this in turn is
proportional to $\tilde m$.
Given the close relation between the order parameter $\tilde m$ and
$W_\tau $ it is not surprising that the effects of the {\it phase transition} described by
$\tilde m$ are so clearly displayed by $P(W_\tau )$.}.

In this Letter we have highlighted the singular behavior of the large fluctuations of a quantity
-- the active work $W_\tau$ done by a tagged particle -- in \textcolor{black}{different}
models \textcolor{black}{representing a large class of} self-propelled particle systems. 
We have shown that\textcolor{black}{, in all cases considered,} a threshold
value $W^\dag _\tau$ exists separating regimes where fluctuations behave in a radically different way.
\textcolor{black}{This has been interpreted as due to a transition -- occurring at the level of fluctuations -- between a {\it gaseous} phase, where the particle internal energy is spent into self-propulsion, and one where this energy supports cluster formation.
An order parameter describing the change has been also identified.
The associated breakdown of the LDP reflects the importance, in terms of probabilistic weight, of clustering-related fluctuations with respect to thermal ones.
For $W_\tau <W^\dag _\tau$, the former are relevant to the large-scale/long-time dynamics,
even for those values of the model parameters for which the whole system is
in a gaseous phase.}

To the best of our knowledge, this is the first evidence of a fluctuation pattern of this kind
in an interacting model of active matter. \textcolor{black}{In this respect, we remark that a singular distribution was also found in a model with  a single active particle in an external field~\cite{seifert2016newjPhy}, suggesting this feature to be generic of self-propelled particles. In addition, the cruciality of interaction among particles in causing the spontaneous breaking of the orientational symmetry makes our results fundamentally different from those obtained both for the solvable cases of transitions at the fluctuating level mentioned above 
\cite{zannetti2014ab,harris2005jstat,szavitnossan2014aa} and in the specific context of active matter.}

Besides the interest of the phenomena described insofar,  this study also shows that a careful 
analysis of non-equilibrium fluctuations may be a sophisticated tool to uncover important dynamical 
properties which would be missed with more conventional analytical methods.  
\textcolor{black}{LDP violations for  $W_\tau < W^\dag _\tau$ enhance
  the probability of the corresponding events,
  possibly with important consequences on specific properties
or functions associated to the work done by active particles.}
This prompts further studies of the fluctuation spectra in
active matter systems, particularly those of biological interest.

Simulations were ran at Bari ReCaS e-Infrastructure funded by MIUR through PON Research and Competitiveness 2007-2013 Call 254 Action I.
FC and GG acknowledge  MIUR for funding (PRIN 2015K7KK8L and PRIN 2012NNRKAF, respectively).

\bibliographystyle{apsrev4-1.bst}
\bibliography{dumbbells-biblio}

\end{document}